\documentclass[footinbib,aps,pra,reprint,superscriptaddress,nopacs]{revtex4-1}

\usepackage{amsmath,amssymb,bbm,bm,graphicx,braket}
\usepackage[absolute]{textpos}
\usepackage{graphicx}
\usepackage{xcolor}
\def\bp{\mathbf{p}}
\def\bbeta{\bm{\beta}}
\def\bx{\mathbf{x}}
\def\by{\mathbf{y}}
\def\bq{\mathbf{q}}

\begin{document}
\title{Efficient compressive and Bayesian characterization of biphoton frequency spectra}

\author{Emma M. Simmerman}
\affiliation{Quantum Information Science Group, Computational Sciences and Engineering Division, Oak Ridge National Laboratory, Oak Ridge, Tennessee 37831, USA}

\author{Hsuan-Hao Lu}
\affiliation{School of Electrical and Computer Engineering and Purdue Quantum Science and Engineering Institute, Purdue University, West Lafayette, Indiana 47907, USA}

\author{Andrew M. Weiner}
\affiliation{School of Electrical and Computer Engineering and Purdue Quantum Science and Engineering Institute, Purdue University, West Lafayette, Indiana 47907, USA}

\author{Joseph M. Lukens}
\email{lukensjm@ornl.gov}
\affiliation{School of Electrical and Computer Engineering and Purdue Quantum Science and Engineering Institute, Purdue University, West Lafayette, Indiana 47907, USA}

\begin{abstract}
Frequency-bin qudits constitute a promising tool for quantum information processing, but their high dimensionality can make for tedious characterization measurements. Here we introduce and compare compressive sensing and Bayesian mean estimation for recovering the spectral correlations of entangled photon pairs. Using a conventional compressive sensing algorithm, we reconstruct joint spectra with up to a 26-fold reduction in measurement time compared to the equivalent raster scan. Applying a custom Bayesian model to the same data, we then additionally realize reliable and consistent quantification of uncertainty. These efficient methods of biphoton characterization should advance our ability to use the high degree of parallelism and complexity afforded by frequency-bin encoding.
\end{abstract}

\maketitle

Given the sparsity of many signals of interest in the real world, compressive sensing (CS)~\cite{candes2008introduction} has emerged as a powerful, general technique for reconstructing a signal from significantly fewer measurements than required by traditional sampling methods. In the context of quantum information, CS has been utilized for quantum state reconstruction~\cite{gross2010quantum, mirhosseini2014compressive, ahn2019adaptive}, process tomography~\cite{shabani2011efficient}, and ghost imaging~\cite{magana2013compressive,morris2015imaging}, and has proven to be an effective tool for efficient characterization of high-dimensional quantum states~\cite{howland2013efficient, montaut2018compressive, schneeloch2019quantifying}, especially in the spatial degree of freedom (DoF) with the aid of digital micromirror devices. 

The approach of Bayesian mean estimation (BME) is more general, in the sense that neither system sparsity nor a particular class of measurements are required for the procedure's validity, though this knowledge can nonetheless be neatly accounted for in the prior distribution. This is an advantage of Bayesian methods, along with return of confidence intervals commensurate with the data gathered~\cite{blume2010optimal}. BME has been explored experimentally, e.g., in single-~\cite{kravtsov2013experimental} and two-qubit polarization quantum state tomography experiments~\cite{PhysRevA.93.012103,williams2017quantum}.

Recently, the frequency DoF has developed into a promising platform for photonic quantum information processing (QIP)~\cite{lukens2017frequency, Kues2019, lu2019quantum}, yet despite the naturally high-dimensional nature of this Hilbert space, CS techniques have yet to be leveraged to characterize spectral properties. And while Bayesian methods have been applied to recover density matrices~\cite{Lu2018b} and mode transformations~\cite{lu2019controlled} in frequency-bin QIP, they have not been explored for extracting high-dimensional biphoton frequency correlations. In this work, we retrieve the frequency correlations of quantum states with CS and BME for the first time.



We consider biphotons generated by spontaneous parametric downconversion, which can exhibit strong frequency anticorrelations and are a common source of information carriers in frequency-domain QIP. Biphoton correlations can be assessed by measuring the joint spectral intensity (JSI). The conventional raster scan method of obtaining the JSI for an $N$-dimensional biphoton Hilbert space ($\sqrt{N}$ dimensions per photon) requires $N$ coincidence measurements between two narrow spectral passbands, one for each photon. One way to improve on this approach is to leverage dispersion in a time-of-flight spectrometer, converting from spectral to temporal correlations and using time-tagging to map out the JSI~\cite{Chen2017,Davis2017,Davis2018,montaut2018compressive}. However, this requires a timing reference to determine absolute---not just relative---frequency, a condition which is not satisfied by, e.g., a free-running continuous-wave--pumped biphoton source. Our approach, 
valid for asynchronous sources, is to perform measurements over many passbands at once. 
We use Fourier-transform pulse shaping ~\cite{Weiner2000} to apply a code of spectral filters to the frequency bins of each photon, with transmission values taken from random binary, random gray-level, or Hadamard codes (selected in random order from the rows of a Hadamard matrix).

In particular, the use of Hadamard codes is well established in classical spectroscopy, where measuring linear combinations of frequency bands can enable significantly higher signal-to-noise ratios (SNRs) in background-limited environments~\cite{nelson1970hadamard}---features which have been explored in measuring correlations of frequency-entangled photons as well~\cite{lukens2013biphoton}, though without recovering the underlying probability distribution. Likewise, the CS and Bayesian methods we consider here enjoy an SNR improvement from measuring many bins at once. CS goes one step further, reducing the total number of measurements needed by exploiting the sparsity anticipated for our biphoton system. BME takes a slightly different perspective on the problem: whereas CS seeks to exceed a threshold of measurements needed to find the underlying signal, BME instead is formulated to return a credible estimate given any collection of measurements. 

%

\begin{figure}[bt!]
\centering
\includegraphics[width=3.4in]{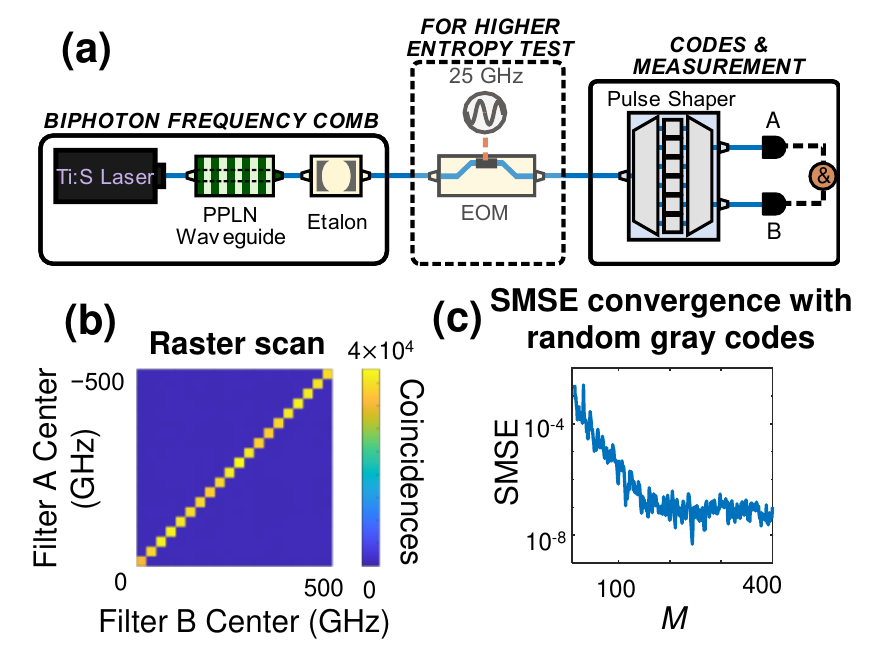}
\caption{(a) Experimental setup. (b) Raster scan of lower entropy state. (c) SMSE number-of-measurement convergence of LASSO reconstructions using random gray value codes.}
\label{fig1}
\end{figure}

The experimental setup is depicted in Fig.~\ref{fig1}(a). Broadband biphotons were generated by pumping a periodically poled lithium niobate (PPLN) ridge waveguide (AdvR) with a continuous-wave 780~nm diode laser under type-0 phase matching. A 25~GHz-spaced etalon was used to produce a biphoton frequency comb. We used a pulse shaper (Finisar) to selectively attenuate 20 energy-anticorrelated frequency bins on either side of the center frequency (limited by the edge of the acceptance bandwidth, giving JSI dimensionality of $N=400$) according to the applied code, while blocking bins outside this range. Signal and idler bins were sent to separate superconducting nanowire single-photon detectors (Quantum Opus) and a time tagger (PicoQuant) ascertained coincidences. We compare reconstructed JSIs obtained from length-20 random and Hadamard codes applied to each photon with 0.5~s integration time to raster scans obtained in 5~s, maintaining the same average singles counts (both random and Hadamard codes average a transmissivity of 50\% over all bins). We first tested a relatively sparse, highly anticorrelated JSI, then added an electro-optic phase modulator (EOM) driven with 25~GHz RF signal to produce spectral sidebands. 

Given the results of an experiment, conventional CS techniques associate the measured coincidences ($y_i$) with a linear function of the joint spectral bins passed in measurement setting $i$ (a length-$N$ vector $\bx_i$ given by the Kronecker product of the individual codes applied to each photon~\cite{howland2013efficient}), weighted by the length-$N$ vector $\bbeta$. Each element of $\bbeta$ is defined as the biphoton flux within a particular bin-pair of the JSI. From this, the $N$-dimensional probability distribution follows as $\bp=\frac{\bbeta}{\sum_i \beta_i}$. For $M$ measurements, $\bbeta$ can be estimated by solving the linear system $\by= A\bbeta$, where $A$ is an $M\times N$ matrix with rows~$\bx_i^T$. 

In order to incorporate knowledge of sparsity, we use the CS least absolute shrinkage and selection operator (LASSO) method \cite{tibshirani1996regression} to find an estimate of $\bbeta$, by solving
\begin{equation}
\bbeta_\mathrm{LASSO} = \min_{\bbeta} \left\{ \frac{1}{2M}\sum_{i=1}^{M}(y_i-\beta_0-\bx_i^T\bbeta)^2+\lambda \sum_{j=1}^{N}|\beta_j| \right\}
\label{lasso}
\end{equation}
with respect to $\bbeta$ and $\beta_0$ (an intercept). The absolute value accounts for LASSO's domain over all real numbers---although as needed in our case, only positive $\beta_i$ are returned.
The first term favors solutions that minimize error with respect to the measurement results, while the second term enforces sparsity. 
We used MATLAB's built-in LASSO algorithm to perform the minimization. In order to select a sensible value for the weight parameter $\lambda$, 
10-fold cross validation was used, which partitions the measurements into training and validation sets and computes the mean squared error (MSE) of each solution against the test data. Multiple $\lambda$ values were tested, and the largest within one standard error of the MSE-minimizing value was chosen.

We first use LASSO to reconstruct a relatively sparse JSI, the raster scan of which appears in Fig.~\ref{fig1}(b). The ratio of diagonal to off-diagonal coincidences is $\sim$40. If nothing is known about the expected form of the JSI, one can assess the number of codes necessary for CS reconstruction by calculating a serial MSE (SMSE) $\frac{1}{N} \sum_{j=1}^{N}(p_j-\overline{p}_j)^{2}$ between each normalized reconstructed JSI $\bp$ (from $M$ codes) and the mean  $\overline{\bp}$ of the previous several ($M-5$ to $M-1$ in our case), continuing to add measurements until the SMSE stabilizes. SMSE LASSO convergence calculated for $M=10$ to $M=N$ is shown in Fig.~\ref{fig1}(c). This approach highlights how one can determine convergence without explicitly comparing the recovered result to a theoretical prediction.

\begin{figure}
\centering
\includegraphics[width=3.4in]{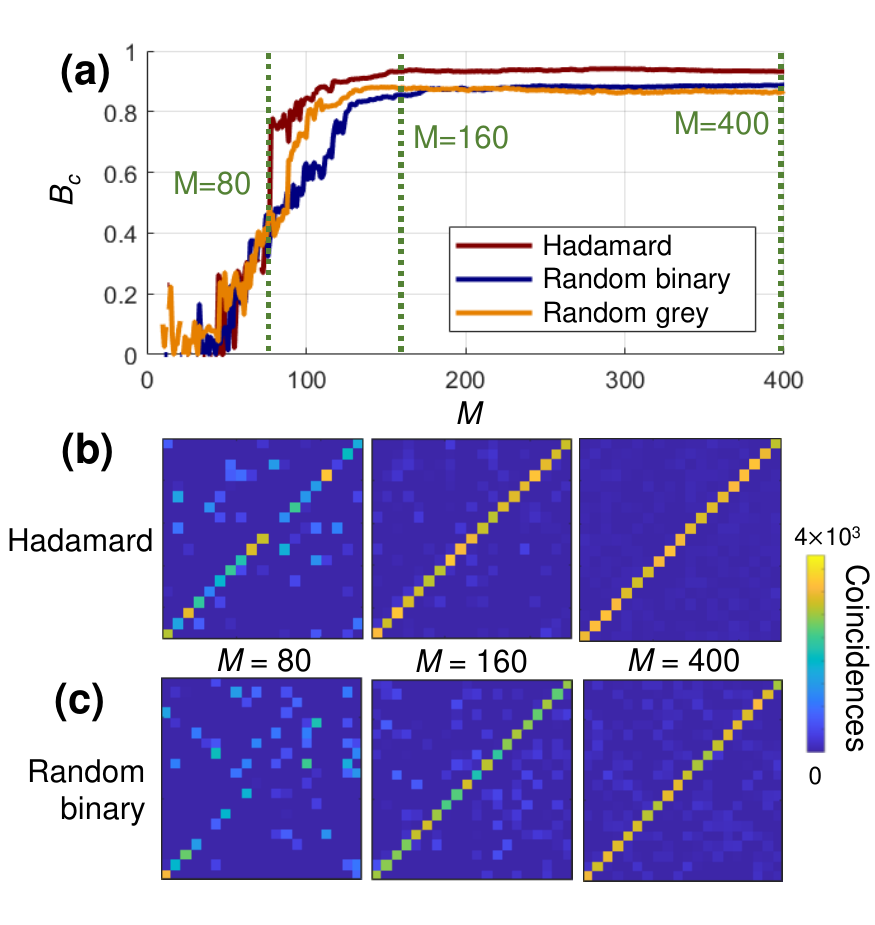}
\caption{LASSO reconstruction. (a) Number-of-measurement convergence. Overlap coefficients ($B_c$) computed with respect to the ideal distribution $\bq$. (b-c) Representative reconstructions using (b) Hadamard and (c) random binary codes.}
\label{fig2}
\end{figure}

If one \emph{can} predict the form of the JSI, however---as in the present case---convergence can also be assessed by comparing the reconstruction at various $M$s to the ideal case with probability vector $\bq$, where, because of the broadband nature of our source, all anti-diagonal bin-pairs are taken to have equal probability and all others are zero. We use the Bhattacharyya coefficient~\cite{Fuchs1999} as a metric of overlap for this purpose, defined as $B_c=\sum_{i=1}^{N}\sqrt{p_i q_i}$. $B_c$ LASSO convergence is shown in Fig.~\ref{fig2}(a). A transition occurs around $M=160$, beyond which additional measurements produce minimal improvement. 
Representative reconstructed JSIs for Hadamard and random binary codes are shown in Fig.~\ref{fig2}(b) and (c), respectively. Each Hadamard sequence has ``1'' in the first column, so that one spectral bin is passed for both signal and idler photons in every measurement. This prevents meaningful extraction of information about the corresponding joint frequency bins in the reconstructed JSI. Consequently, these bins were removed in the Hadamard reconstructions below, making their JSIs 19$\times$19.
(Incidentally, conventional Hadamard spectroscopy omits the first row and column of a Hadamard matrix when defining its code sequences for this reason~\cite{nelson1970hadamard}.)

In order to compare the number of measurements required for convergence with that anticipated from CS theory, we can quantify the JSI sparsity by the effective number of states, $K=2^{H(\bp)}$, with $H(\bp)$ the information entropy $H(\bp)=-\sum_{j=1}^{N}p_j\log_{2}{p_j}$. Experimentally, this number falls between that of an ideal uniform anticorrelated JSI ($K=20$) and the maximum uncorrelated case of $K=N=400$. The shoulder at $M=160$ [Fig.~\ref{fig2}(a)] corresponds to more measurements than the $\sim$60 expected for a $K=20$ sparse system, based on the scaling $K\ln\frac{N}{K}$~\cite{candes2008introduction, howland2013efficient}. However, computing the effective dimensionality $K$ from the raster scan itself gives $K=97$, an increase which can be attributed to the off-diagonal background counts. Interestingly, $K=97$ predicts a required number of measurements of $\sim$140, much closer to our observations. 
The 10-fold improvement in the acquisition time (at the same average single-photon flux), combined with the advantage of fewer measurements, results in a 26-fold reduction in the measurement time compared to the raster scan. Each reconstruction (including cross-validation) took less than two minutes on a personal laptop.

As Eq.~(\ref{lasso}) corresponds to a convex optimization problem, it admits highly efficient numerical solutions; this ease of computation is an advantage of CS methods in general, but it comes at the cost of ambiguity in the level of sparsity enforcement (signified by the parameter $\lambda$ in the case of LASSO). Even when comparing different possible values, as by cross-validation, the final choice of $\lambda$ is ultimately subjective. BME, on the other hand, ``offers one answer to a well-posed problem''~\cite{mackay2003information}. That is, the estimator's initial model of the system (which can be made as uninformed as necessary to reflect actual knowledge) together with the data, uniquely determine the posterior distribution. Therefore, as long as the probability model can be justified by physical principles and the resulting distribution is adequately sampled, the result of BME is unambiguous.

Specifically, Bayes' theorem gives the posterior probability distribution of parameters representing the state of a system as the normalized product of the likelihood function and prior distribution \cite{mackay2003information}. The likelihood is the probability of observing the collected data (the coincidence record $\by$) given a possible state of the system, according to some model, while the prior distribution represents initial knowledge of the state of the system. Again, we seek the vector $\bbeta$ (biphoton fluxes), but we can now infer the underlying probability distribution $\bp=\frac{\bbeta}{\sum_{i}\beta_{i}}$ directly, by explicitly introducing a to-be-determined scaling parameter $C$ such that $\bbeta=C\bp$. 
Then, for the same $\bx_i$ codes as in Eq.~(\ref{lasso}), we can take as our likelihood
\begin{equation}
\label{likelihood}
    \mathcal{P}(\by|\bp,C)\propto\prod_{i=1}^{N}e^{-C \bx_i^T \bp}\left(C \bx_i^T \bp\right)^{y_i},
\end{equation}
which models each observation according to a Poisson distribution of mean $C\bx_i^T \bp$. We assign a prior on $C$ as a normal distribution of mean $C_0$ and standard deviation $0.1C_0$, where $C_0$ is set initially by averaging the coincidences (we found the chosen variance sufficient for a fully noninformative prior over all feasible values of $C$). 
For $\bp$'s prior, we draw from an $N$-dimensional Dirichlet distribution $\mathrm{Dir}(\alpha)$ to enforce normalization and nonnegativity. The parameter $\alpha$ can be set to favor sparser solutions, but interestingly, we found this to have minimal impact on the result. 

\begin{figure}[t!]
\centering
\includegraphics[width=3.4in]{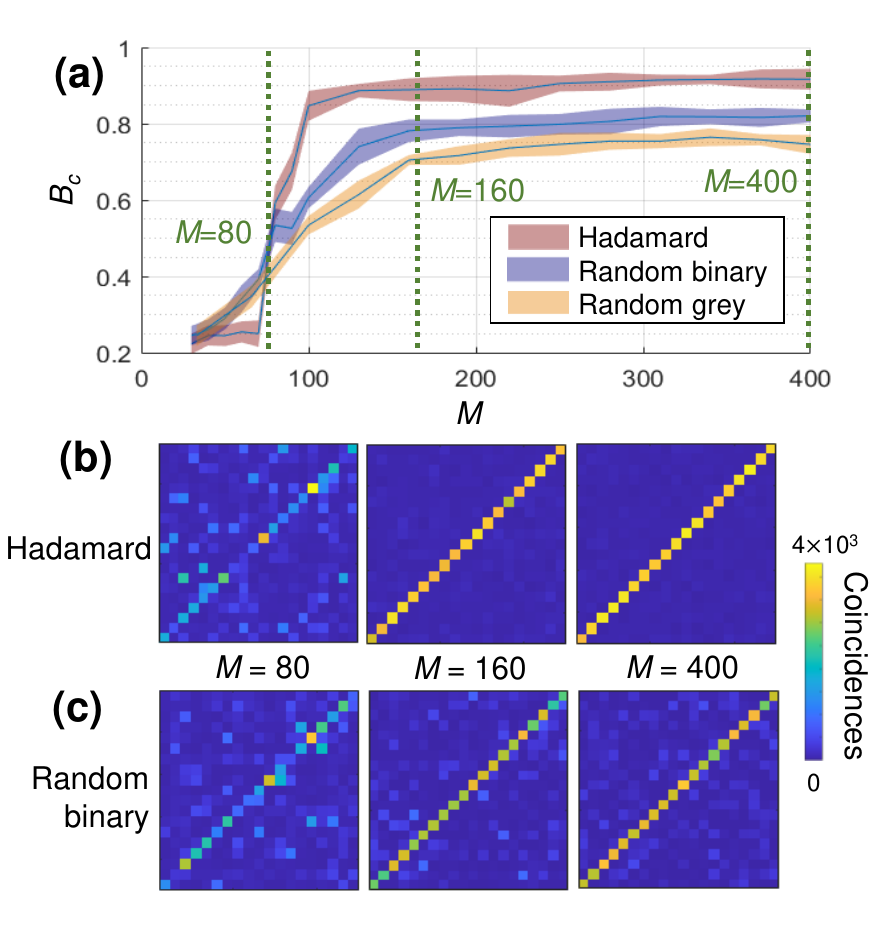}
\caption{Bayesian reconstruction. (a) Number-of-measurement convergence. Overlap coefficients ($B_c$) computed with respect to the ideal distribution $\bq$. (b-c) Representative reconstructions using (b) Hadamard and (c) random binary codes.}
\label{fig3}
\end{figure}

Given the collected data and an appropriate prior distribution $\mathcal{P}(\bp,C)$, the BME result for any function $\phi(\bp,C)$ is given by the mean over the posterior distribution $\mathcal{P}(\bp,C|\by)\propto \mathcal{P}(\by|\bp,C) \mathcal{P} (\bp,C)$, i.e., $\langle\phi\rangle = \int d\bp dC\, \mathcal{P}(\bp,C|\by) \phi(\bp,C)$. Occasionally such an integral may be solved analytically~\cite{williams2017quantum}, but more commonly a numerical sampling method is required. Here we use a preconditioned Crank--Nicolson Metropolis--Hastings algorithm (recently introduced in the context of quantum state tomography~\cite{Law2020}), a Markov-chain Monte Carlo Method which iteratively generates samples from the posterior distribution, accepting new samples with a probability based on the evaluated likelihood-prior product at each point, and the proposal density function. In this way the posterior distribution is sampled more heavily around local maxima, while allowing for jumps to lower-probability regions, effectively sampling the entire space.


To ensure convergence, we increased the number of samples in the Markov chain until $B_c$ values no longer varied. 
We found that tuning the $\alpha$ parameter in the prior to favor sparse solutions ($\alpha<1$) made no considerable difference in reconstructions, confirming one prominent advantage of Bayesian methods---that the form of the prior becomes irrelevant if there is a sufficient amount of data. $B_c$ BME convergence is shown in Fig.~\ref{fig3}(a), along with standard deviation error bounds. Reconstructions with $M\sim160$ took about 12 minutes each on the same laptop as used previously, considerably longer than LASSO. This is the chief disadvantage of Bayesian methods, making the SMSE approach to number-of-measurement convergence nonviable here. The counteracting advantage, however, is the return of appropriate error bounds for the legitimacy of each reconstruction. For BME, Hadamard codes produced notably better reconstructions than random codes, and number-of-measurement convergence occurred around $M$=160, as with LASSO, and with similar $B_c$ values at convergence. Thus BME affords the same $\sim$26-fold reduction in \emph{experimental} measurement time (though of course with significantly longer numerical analysis). Representative reconstructed JSIs for Hadamard and random binary codes are shown in Fig.~\ref{fig2}(b) and (c), respectively, for $M$= 80, 160, and 400.

\begin{figure}
\centering
\includegraphics[width=3.3in]{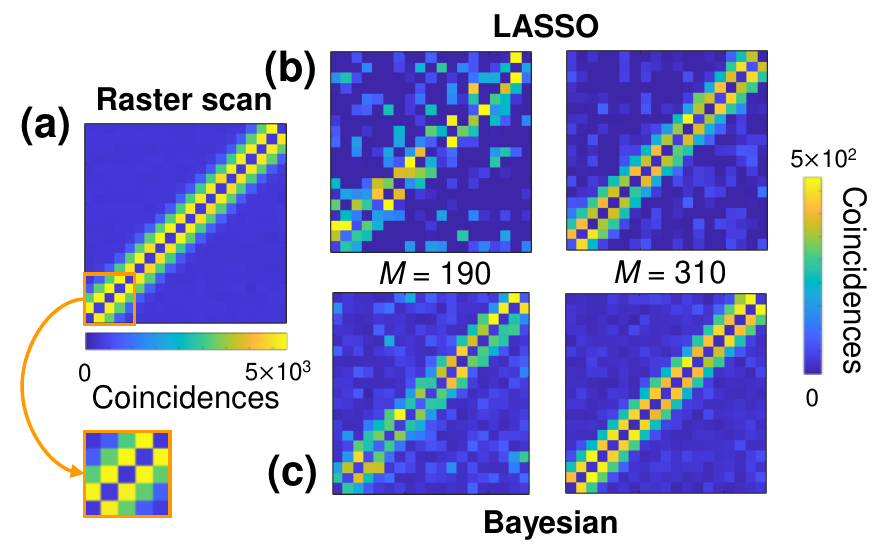}
\caption{Higher entropy state. (a) JSI raster scan. (b) LASSO and (c) BME reconstructions using Hadamard codes.}
\label{fig4}
\end{figure}

With the introduction of an EOM, more complicated JSIs can be explored as well~\cite{IMANYwalk}, due to the presence of additional sidebands. Increasing the phase modulation amplitude to split the original diagonal into two main peaks, we measured the JSI with raster scan shown in Fig.~\ref{fig4}(a), with a zoom-in highlighting the split interference pattern. At the chosen level of modulation, the effective number of states for an ideal input state is expected to reach $K=89$, nearly a quarter of the total dimensionality---and, due to off-diagonal background, the number of states computed from the raster scan is even more, at $K=205$---so the applicability of sparsity-based CS is questionable in this case. Nevertheless, though we did not observe a clear $B_c$ convergence point, we did find reasonable reconstruction with $M\sim$300 Hadamard codes. Figure~\ref{fig4}(b) and (c) show representative LASSO and BME reconstructions, respectively. With a more complex JSI, BME visibly outperforms LASSO, especially at reconstructions with fewer measurements (as expected). While reconstructions are clearly noisier than the raster scan, the total measurement time is over 16 times shorter, indicating a practical advantage for rapid measurements in photon-starved environments. 

In future work, thresholding~\cite{howland2013efficient} could be applied to further enhance the contrast of the recovered JSIs. Interestingly, we anticipate advantages from our Bayesian approach as well. Whereas selecting a particular threshold in conventional CS is, like choosing $\lambda$, somewhat ad hoc, it should be possible to incorporate background into the likelihood [Eq.~(\ref{likelihood})] using a physically motivated model for accidentals---similar to the methods developed in Ref.~\cite{lu2019controlled}. In this way, our Bayesian approach to JSI reconstruction should offer a flexible framework which can be specialized to a variety of situations in biphoton characterization.

\textbf{Funding.} U.S. Department of Energy (DOE), Office of Science:  Office of Workforce Development for Teachers and Scientists (WDTS) Science Undergraduate Laboratory Internship program; and Office of Advanced  Scientific Computing Research, Early Career Research program.

\textbf{Acknowledgments.} We thank AdvR, Inc., for loaning the PPLN ridge waveguide and P. Lougovski for discussions. This research was performed in part at Oak Ridge National Laboratory, managed by UT-Battelle, LLC, for the U.S. Department of Energy under contract no. DE-AC05-00OR22725.

\end{document}